\documentstyle[12pt,sw20lart]{article}

\input tcilatex
\begin{document}

\title{{\rm MODERN STATUS OF NEUTRINO EXPERIMENTS AT THE UNDERGROUND NEUTRINO
LABORATORY OF KURCHATOV INSTITUTE NEAR KRASNOYARSK NUCLEAR REACTOR}}
\author{{\rm Yu.V.Kozlov, S.V.Khalturtsev, I.N.Machulin,} \and {\rm \
A.V.Martemyanov, V.P.Martemyanov, A.A.Sabelnikov,} \and {\rm S.V.Sukhotin,
V.G.Tarasenkov, E.V.Turbin, V.N.Vyrodov } \\
${\it Russian\ Reseacrch\ Center\ "Kurchatov\ Institute",}$\\
${\it \ 123182Moscow,\ Russia}$}
\maketitle

\begin{abstract}
{\rm The investigation of antineutrino-deuteron interaction at Krasnoyarsk
reactor are discussed.\ The characteristics of the installation
''Deuteron'', present results and perspectives of Krasnoyarsk neutrino
laboratory are presented.}
\end{abstract}

\section{Introduction}

\ \quad {\rm This report presents the preliminary results of the
experiments, which is in progress at the neutrino underground laboratory
near Krasnoyarsk reactor.}

{\rm Reactor antineutrinos (}$\tilde \nu _e${\rm ) interaction with deuteron
comes by two channels:\centerline{and}}

\begin{eqnarray}
\tilde \nu _e+d &\rightarrow &p+n+\tilde \nu _e^{\prime }\qquad {\rm (NCD\
-\ Neutral\ Current\ on\ Deuteron)} \\
\tilde \nu _e+d &\rightarrow &n+n+e^{+}\qquad {\rm (CCD\ -\ Charged\
Current\ on\ Deuteron)}
\end{eqnarray}
$\ $

{\rm For the first time an investigation of these two reactions was proposed
by Yu.Gaponov and I.Tyutin \cite{Gap}, where they also calculated the cross
sections for both reactions.}

{\rm The studying of these reactions can give information about:}

{\rm a) weak constants for charged and neutral currents;}

{\rm b) a length of neutron - neutron scattering;}

{\rm c) neutrino oscillations;}

{\rm The investigation of this interaction with reactor antineutrinos has
advantages compared with accelerator neutrinos because of reactor
antineutrinos interact with deuteron near a threshold of these reactions .}

{\rm Measuring of the cross section in CCD channel permits to get a value of
length of neutron - neutron scattering a}$_{nn}${\rm \ in S channel without
any strong interaction corrections, because neutrons of this reaction final
step are in S stage. A table number 1 presents a dependence of a}$_{nn}${\rm %
\ on \TEXTsymbol{<}}$\sigma _{cc}${\rm \TEXTsymbol{>}:}

\begin{center}
\begin{tabular}{|cc|}
\multicolumn{2}{c}{\rm Table 1} \\ \hline
\multicolumn{1}{|c|}{{\rm a}$_{nn}${\rm (S), fm}} & {\rm \TEXTsymbol{<}}$%
\sigma _{cc}${\rm \TEXTsymbol{>}x10}$^{44}{\rm ,cm}^2{\rm /fiss}$ \\ \hline
\multicolumn{1}{|c|}{\rm -16.6} & {\rm 1.077} \\ 
\multicolumn{1}{|c|}{\rm -17.0} & {\rm 1.084} \\ 
\multicolumn{1}{|c|}{\rm -18.5} & {\rm 1.112} \\ 
\multicolumn{1}{|c|}{{\rm -23.7(=a}$_{np}{\rm )}$} & {\rm 1.179} \\ \hline
\end{tabular}
\end{center}

{\rm Existing experimental data for a}$_{nn}${\rm \ have more less accuracy
than for }${\rm a}_{np}${\rm \ and badly conform one with others. Average
meaning of experimental value for a}$_{nn}${\rm (S) is (-16.6}${\rm \pm
0.6)fm\ and\ the\ last\ result\ is\ (-18.5\pm 0.5)fm,\ }$

${\rm whereas\ {}a}_{np}{\rm =(-23.715)fm.}$

\section{{\bf Three experimental results obtained with reactor antineutrinos
for today}{\rm .}}

\ \quad {\rm Results of previous reactor experiments on studying
antineutrino- deuteron interaction are shown in table 2:}

\begin{center}
\begin{tabular}{|p{2cm}p{6cm}p{4cm}|}
\multicolumn{3}{c}{\rm Table 2} \\ \hline
\multicolumn{1}{|p{2cm}|}{Savannah} & \multicolumn{1}{p{6cm}|}{$\sigma
^{NCD}=(3.8\pm 0.9)\times 10^{-45}cm^{2}/\nu _{e}$} & $\sigma
_{exp}^{NCD}/\sigma _{theor}^{NCD}=0.8\pm 0.2$ \\ 
\multicolumn{1}{|p{2cm}|}{River \cite{Pas}:} & \multicolumn{1}{p{6cm}|}{$%
\sigma ^{CCD}=(1.5\pm 0.4)\times 10^{-45}cm^{2}/\nu _{e}$} & $\sigma
_{exp}^{CCD}/\sigma _{theor}^{CCD}=0.7\pm 0.2$ \\ 
\multicolumn{1}{|p{2cm}|}{} & \multicolumn{1}{p{6cm}|}{$\sigma
_{exp}^{CCD}/\sigma _{exp}^{NCD}=0.40\pm 0.14$} & $\sigma
_{theor}^{CCD}/\sigma _{theor}^{NCD}=0.353^{*}$ \\ \hline
\multicolumn{1}{|p{2cm}|}{Krasno-} & \multicolumn{1}{p{6cm}|}{$\sigma
^{NCD}=(3.0\pm 1.0)\times 10^{-44}cm^{2}/fis.^{235}U$} & $\sigma
_{exp}^{NCD}/\sigma _{theor}^{NCD}=0.95\pm 0.33^{*}$ \\ 
\multicolumn{1}{|p{2cm}|}{yarsk \cite{Koz}:} & \multicolumn{1}{p{6cm}|}{$%
\sigma ^{CCD}=(1.1\pm 0.2)\times 10^{-44}cm^{2}/fis.^{235}U$} & $\sigma
_{exp}^{CCD}/\sigma _{theor}^{CCD}=0.98\pm 0.18^{*}$ \\ 
\multicolumn{1}{|p{2cm}|}{} & \multicolumn{1}{p{6cm}|}{$\sigma
_{exp}^{CCD}/\sigma _{exp}^{NCD}=0.37\pm 0.14$} & $\sigma
_{theor}^{CCD}/\sigma _{theor}^{NCD}=0.353^{*}$ \\ \hline
\multicolumn{1}{|p{2cm}|}{Rovno \cite{Ver}:} & \multicolumn{1}{p{6cm}|}{$%
\sigma ^{NCD}=[2.71\pm 0.46(stat)\pm 0.11(sys)]\times $} & $\sigma
_{exp}^{NCD}/\sigma _{theor}^{NCD}=0.92\pm 0.18$ \\ 
\multicolumn{1}{|p{2cm}|}{} & \multicolumn{1}{p{6cm}|}{$\times
10^{-44}cm^{2}/fis.PWR-440$} &  \\ 
\multicolumn{1}{|p{2cm}|}{} & \multicolumn{1}{p{6cm}|}{$\sigma
^{CCD}=[1.17\pm 0.14(stat)\pm 0.07(sys)]\times $} & $\sigma
_{exp}^{CCD}/\sigma _{theor}^{CCD}=0.37\pm 0.08$ \\ 
\multicolumn{1}{|p{2cm}|}{} & \multicolumn{1}{p{6cm}|}{$\times
10^{-44}cm^{2}/fis.PWR-440$} &  \\ 
\multicolumn{1}{|p{2cm}|}{} & \multicolumn{1}{p{6cm}|}{$\sigma
_{exp}^{CCD}/\sigma _{exp}^{NCD}=0.43\pm 0.10$} & $\sigma
_{theor}^{CCD}/\sigma _{theor}^{NCD}=0.37\pm 0.08$ \\ \hline
\end{tabular}
\end{center}

$\sigma _{theor}^{NCD}=3.124\times 10^{-44}cm^2/fis.^{235}U^{*}${\rm \ }$%
\sigma _{theor}^{CCD}=1.11\times 10^{-44}cm^2/fis.^{235}U^{*}${\rm \footnote{%
{\rm * -- here it is the theoretical values of cross sections taken from the
work of Gaponov Yu.V. and Vladimirov D.M. for Schreckenbach K. reactor
antineutrino spectrum\cite{Gap2}.}}}

\section{\bf Detector.}

\quad \ {\rm The modernized detector ''Deuteron'' (Fig.1) is situated at the
underground laboratory (600 meters of water equivalent) at the distance of
34.0 m from the reactor, flux of neutrino is about a few units to 10}$^{12}\ 
\widetilde{\nu }${\rm /cm}$^2${\rm . The target }$\sqrt{\sqrt{\sqrt{}}}${\rm %
volume is 513 liters of }$D_2O${\rm \ (or }$H_2O${\rm ) placed in the
stainless tank, which is surrounded by 30 cm of Teflon for neutron
reflection, 0.1 cm of Cd, 8cm of steel shots, 20cm. of graphite and 16cm of
boron polyethylene (}$CH_2+3\%B${\rm ) for gamma and neutron shielding. The
whole installation is pierced to make 169 holes (81 holes pass through the
tank and Teflon, the others through the Teflon only). These holes house 169
proportional }$^3He${\rm \ neutron counters with a reduced intrinsic alpha
background. These counters are used for the neutron registration. They are
located in the square lattice with a square side 10 cm. The active shielding
covers the main assembly, against the cosmic muons.}

{\rm The neutron counters used in the experiment can register only neutrons,
so this detector is a detector of an integral type.}

{\rm The counter consist of a stainless steel tube having length of 1 m and
31 mm in diameter with wall thickness of 0.5 mm. A 20 micron wire is
extended along the counter. The wire is made from tungsten, covered by gold.
Counter inner surface is covered by 60 microns of Teflon layer to reduce the
natural alpha-background from a stainless steel wall, and than the Teflon
layer is covered by 2 microns of pure copper layer to keep the counter able
to work. The counter is filled with a mixture of 4 KPa }$^3${\rm He and 4 KPa%
}$^{40}${\rm Ar gases.}

{\rm The main characteristics of the detector are presented in table 3:}

\begin{center}
\begin{tabular}{|p{7.5cm}|l|l|}
\multicolumn{3}{c|}{\rm Table 3} \\ \hline
{\rm Parameters\TEXTsymbol{\backslash}Target} & {\rm Light water} & {\rm %
Heavy water} \\ \hline
{\rm An efficiency of one neutron registration by only tank counters} & {\rm %
(27.5}${\rm \pm 0.3)\%}$ & {\rm (56.0}${\rm \pm 0.7)\%}$ \\ \hline
{\rm An efficiency of double neutron registration by all counters} & {\rm %
(9.9}${\rm \pm 0.1)\%}$ & {\rm (41.6}${\rm \pm 0.4)\%}$ \\ \hline
{\rm A neutron life time} & {\rm (138}${\rm \pm 2)\mu \sec }$ & {\rm (203}$%
{\rm \pm 2)\mu \sec }$ \\ \hline
\end{tabular}
\end{center}

{\rm The efficiencies of the neutron registration for all neutrino reactions
(CCP, NCD and CCD) were obtained by means of Monte-Carlo (MC) calculations.
The reliability of the MC programs was checked by comparing the MC
calculations with the corresponding experimental measurements with a }$%
^{252} ${\rm Cf neutron source. The divergency between the calculations and
the experimental data for all checks was within 1\%.}

\begin{center}
\FRAME{ftbpF}{377.1875pt}{166.375pt}{0pt}{}{}{Figure }{\special{language
"Scientific Word";type "GRAPHIC";display "USEDEF";valid_file "T";width
377.1875pt;height 166.375pt;depth 0pt;original-width
752.4375pt;original-height 285.0625pt;cropleft "0";croptop "1";cropright
"1";cropbottom "0";tempfilename 'neutr1.gif';tempfile-properties "XP";}}
\end{center}

\section{Status of experimental work with the detector.}

\ \quad {\rm To get more information about characteristics of the detector
and improve of limits on some parameters of neutrino oscillations the
detector at first was filled by light water (H}$_2${\rm O) (at the end of
1995) and then by heavy water (at the end of 1996). The measurements with
heavy water (D}$_2${\rm O) have been started at the beginning of 1997 and
are in progress now.}

\section{The results for the inverse beta-decay on proton (CCP reaction).}

\quad \ {\rm The detector exposure with the light water (H}$_{2}${\rm O)
continued for 115 }$\times 10^{5}${\rm \ sec. The results of inverse beta-
decay reaction on protons (}$\widetilde{\nu }+p\rightarrow n+e^{-}{\rm )}$%
{\rm (CCP reaction)(events per day) are shown in table 4. All measurements
were divided on 4 sets because of different background conditions.}

\begin{center}
\begin{tabular}{|p{2cm}|p{3cm}|p{3cm}|p{3cm}|}
\multicolumn{4}{c|}{Table 4} \\ \hline
{\rm Group of measurements} & {\rm Reactor switched on (ON)} & {\rm Reactor
switched off (OFF)} & {\rm Effect (ON\_OFF)} \\ \hline
{\rm Set1} & {\rm 348.6}${\rm \pm 3.9}$ & {\rm 174.0}${\rm \pm 6.5}$ & {\rm %
174.6}${\rm \pm 6.9}$ \\ \hline
{\rm Set2} & {\rm 341.7}${\rm \pm }${\rm 3.1} & {\rm 177.0}${\rm \pm 5.9}$ & 
{\rm 164.8}${\rm \pm 6.7}$ \\ \hline
{\rm Set3} & {\rm 329.5}${\rm \pm 3.4}$ & {\rm 162.3}${\rm \pm 4.9}$ & {\rm %
169.5}${\rm \pm 6.0}$ \\ \hline
{\rm Set4} & {\rm 327.5}${\rm \pm 4.2}$ & {\rm 146.5}${\rm \pm 4.8}$ & {\rm %
180.9}${\rm \pm 6.3}$ \\ \hline
\end{tabular}

{\rm For nominal reactor power }$N_{eff}=(177.2\pm 3.3)~events/10^{5}~sec,$
\end{center}

{\rm which corresponds the value of the CCP cross-section: } 
\begin{equation}
\sigma _{exp}=(6.39\times 10^{-43}\pm 3.0\%)cm^2/fission~^{235}U
\end{equation}

{\rm This result is in a good agreement with the theoretical cross-section
(V-A theory). Their ratio is: } 
\begin{equation}
R=\frac{\sigma _{exp}}{\sigma _{V-A}}(^{235}U)=1.00\pm 0.04\ \qquad (68\%CL)
\end{equation}

{\rm This experiment gives no indications to the neutrino oscillations, so
only the following limitation parameters can be obtained at 90\%CL: } 
\begin{equation}
\Delta m_{1,2}^2\le 0.016eV^2~~~for~the~\sin ^2(2\Theta )=1
\end{equation}
\begin{equation}
\sin ^2(2\Theta )\le 0.09~~~~for~the~\Delta m_{1,2}^2\ge 1eV^2
\end{equation}

{\rm Combining this limits with the results from previous experiments at
Krasnoyarsk reactor the summary limitations on the neutrino oscillation
parameters are presented on figure 2 with the limitations from Bugey \cite
{Ach}, Gosgen\cite{Gos} limitation from Chooz\cite{Cho} experiment and zone
permitted for }$\nu _{e}\leftrightarrow \nu _{\mu }\ ${\rm oscillation
represented by Kamiokande \cite{Fuk} groups.}

\FRAME{ftbpFU}{323.1875pt}{323.0625pt}{0pt}{\Qcb{The 90\% exclussion plots,
compared with the KAMIOKANDE allowed region.}}{}{Figure 2}{\special{language
"Scientific Word";type "GRAPHIC";maintain-aspect-ratio TRUE;display
"USEDEF";valid_file "T";width 323.1875pt;height 323.0625pt;depth
0pt;original-width 0pt;original-height 0pt;cropleft "0";croptop
"1";cropright "1";cropbottom "0";tempfilename
'Neutr2.gif';tempfile-properties "XP";}}

\smallskip

\section{Status of measuring of antineutrino - deuteron interaction.}

\ \quad {\rm From the beginning of 1997 there were measured six sets with
reactor ON (about 211}$\times 10^{5}\sec ${\rm \ ) and six sets with reactor
OFF ( about 62}$\times 10^{5}\sec ${\rm \ ). Preliminary results ( events
per day) were placed in table 5:}

\begin{center}
\begin{tabular}{|p{2cm}|p{3cm}|c|c|c|}
\multicolumn{5}{c|}{\rm Table5} \\ \hline
{\rm Number of SET} & {\rm Reaction (neutron multiplicity)} & {\rm Reactor ON%
} & {\rm Reactor OFF} & {\rm ON - OFF} \\ \hline
{\rm 1} & {\rm NCD(1)} & {\rm 296.6}$\pm ${\rm 3.0} & {\rm 277.2}$\pm ${\rm %
4.2} & {\rm 19.3}$\pm ${\rm 5.2} \\ \hline
& {\rm CCD(2)} & {\rm 22.1}$\pm ${\rm 0.8} & {\rm 17.8}$\pm ${\rm 1.1} & 
{\rm 4.3}$\pm ${\rm 1.3} \\ \hline
{\rm 2} & {\rm NCD(1)} & {\rm 291.2}$\pm ${\rm 3.0} & 27{\rm 3.4}$\pm ${\rm %
4.8} & {\rm 17.8}$\pm ${\rm 5.7} \\ \hline
& {\rm CCD(2)} & {\rm 22.2}$\pm ${\rm 0.8} & {\rm 17.8}$\pm ${\rm 1.2} & 
{\rm 4.4}$\pm ${\rm 1.5} \\ \hline
{\rm 3} & {\rm NCD(1)} & {\rm 289.9}$\pm ${\rm 3.1} & {\rm 279.7}$\pm ${\rm %
6.4} & {\rm 10.2}$\pm ${\rm 7.1} \\ \hline
& {\rm CCD(2)} & {\rm 20.0}$\pm ${\rm 0.8} & {\rm 16.7}$\pm ${\rm 1.6} & 
{\rm 3.3}$\pm ${\rm 1.8} \\ \hline
4 & {\rm NCD(1)} & 415.3$\pm 2.9$ & 389.5$\pm 4.1$ & 25.8$\pm 5.0$ \\ \hline
& {\rm CCD(2)} & 21.5$\pm 0.7$ & 18.0$\pm 0.9$ & 3.5$\pm 1.1$ \\ \hline
5 & {\rm NCD(1)} & 412.4$\pm 2.6$ & 393.8$\pm 6.2$ & 18.6$\pm 6.7$ \\ \hline
& {\rm CCD(2)} & 20.5$\pm 0.6$ & 17.7$\pm 1.3$ & 2.8$\pm 1.4$ \\ \hline
6 & {\rm NCD(1)} & 406.0$\pm 3.3$ & 372.2$\pm 9.3$ & 33.7$\pm 9.9$ \\ \hline
& {\rm CCD(2)} & 19.6$\pm 0.7$ & 14.0$\pm 1.8$ & 5.6$\pm 2.0$ \\ \hline
{\rm Sum} & {\rm NCD(1)} &  &  & {\rm 20.3}$\pm ${\rm 2.5} \\ \hline
{\rm result} & {\rm CCD(2)} &  &  & {\rm 3.8}$\pm ${\rm 0.6} \\ \hline
\end{tabular}
\end{center}

{\rm Processing these results gives:}

\begin{center}
\begin{tabular}{l}
$\sigma ^{NCD}=(3.3\pm 0.4)\times 10^{-44}cm^{2}/fis.U^{235}$ \\ 
$\sigma ^{CCD}=(1.2\pm 0.2)\times 10^{-44}cm^{2}/fis.U^{235}$ \\ 
$\sigma ^{NCD}/\sigma ^{CCD}=0.31\pm 0.07$%
\end{tabular}

{\rm Summering latest results with previous Krasnoyarsk group experiment
we've got:}

\begin{tabular}{l}
$\sigma ^{NCD}=(3.2\pm 0.4)\times 10^{-44}cm^{2}/fis.U^{235}$ \\ 
$\sigma ^{CCD}=(1.2\pm 0.1)\times 10^{-44}cm^{2}/fis.U^{235}$ \\ 
$\sigma ^{NCD}/\sigma ^{CCD}=0.36\pm 0.06$%
\end{tabular}
\end{center}

{\rm The experiment is in progress and we hope to get statistical accuracy
in cross sections of the reactions \symbol{126} 5\%.}

\section{\protect\smallskip New neutrino laboratory near Krasnoyarsk nuclear
reactor}

{\rm A new laboratory hall situated at less distance from the reactor have
been created, a neutrino flux will be 3.5 times more there than at the
previous laboratory hall and it will be about 10}$^{13}\widetilde{\nu }/$cm$%
^{2}$/sec.{\rm \ A scheme of this laboratory is presented by fig.3:}

\FRAME{ftbpFU}{378.3125pt}{284.8125pt}{0pt}{\Qcb{A scheme of new Krasnoyarsk
newtrino laboratory hall.}}{}{Figure 3}{\special{language "Scientific
Word";type "GRAPHIC";maintain-aspect-ratio TRUE;display "USEDEF";valid_file
"T";width 378.3125pt;height 284.8125pt;depth 0pt;original-width
754.3125pt;original-height 284.5625pt;cropleft "0";croptop "1";cropright
"1";cropbottom "0";tempfilename 'Neutr3.gif';tempfile-properties "XNP";}}

\smallskip It is planned to use this new laboratory hall for searching for
neutrino magnetic moment experiment.

\section{Acknowledgments}

\quad \ {\rm We} {\rm should like to thank the staff of Krasnoyarsk reactor
for constant help, Ac. S.T.Belyaev and Dr. Yu.V.Gaponov for very useful
discussions. }This work is supported by RFBR grants NN 96-15-96640 and
98-02-16313.

\end{document}